\documentstyle[wstwocl]{article}
\begin{document}

\bibliographystyle{unsrt}    

\newcommand{\st}{\scriptstyle}
\newcommand{\sst}{\scriptscriptstyle}
\newcommand{\mco}{\multicolumn}
\newcommand{\epp}{\epsilon^{\prime}}
\newcommand{\vep}{\varepsilon}
\newcommand{\ra}{\rightarrow}
\newcommand{\ppg}{\pi^+\pi^-\gamma}
\newcommand{\vp}{{\bf p}}
\newcommand{\ko}{K^0}
\newcommand{\kb}{\bar{K^0}}
\newcommand{\al}{\alpha}
\newcommand{\ab}{\bar{\alpha}}
\def\be{\begin{equation}}
\def\ee{\end{equation}}
\def\bea{\begin{eqnarray}}
\def\eea{\end{eqnarray}}
\def\CPbar{\hbox{{\rm CP}\hskip-1.80em{/}}}

\def\ap#1#2#3   {{\em Ann. Phys. (NY)} {\bf#1} (#2) #3.}
\def\apj#1#2#3  {{\em Astrophys. J.} {\bf#1} (#2) #3.}
\def\apjl#1#2#3 {{\em Astrophys. J. Lett.} {\bf#1} (#2) #3.}
\def\app#1#2#3  {{\em Acta. Phys. Pol.} {\bf#1} (#2) #3.}
\def\ar#1#2#3   {{\em Ann. Rev. Nucl. Part. Sci.} {\bf#1} (#2) #3.}
\def\cpc#1#2#3  {{\em Computer Phys. Comm.} {\bf#1} (#2) #3.}
\def\err#1#2#3  {{\it Erratum} {\bf#1} (#2) #3.}
\def\ib#1#2#3   {{\it ibid.} {\bf#1} (#2) #3.}
\def\jmp#1#2#3  {{\em J. Math. Phys.} {\bf#1} (#2) #3.}
\def\ijmp#1#2#3 {{\em Int. J. Mod. Phys.} {\bf#1} (#2) #3.}
\def\jetp#1#2#3 {{\em JETP Lett.} {\bf#1} (#2) #3.}
\def\jpg#1#2#3  {{\em J. Phys. G.} {\bf#1} (#2) #3.}
\def\mpl#1#2#3  {{\em Mod. Phys. Lett.} {\bf#1} (#2) #3.}
\def\nat#1#2#3  {{\em Nature (London)} {\bf#1} (#2) #3.}
\def\nc#1#2#3   {{\em Nuovo Cim.} {\bf#1} (#2) #3.}
\def\nim#1#2#3  {{\em Nucl. Instr. Meth.} {\bf#1} (#2) #3.}
\def\np#1#2#3   {{\em Nucl. Phys.} {\bf#1} (#2) #3.}
\def\pcps#1#2#3 {{\em Proc. Cam. Phil. Soc.} {\bf#1} (#2) #3.}
\def\pl#1#2#3   {{\em Phys. Lett.} {\bf#1} (#2) #3.}
\def\prep#1#2#3 {{\em Phys. Rep.} {\bf#1} (#2) #3.}
\def\prev#1#2#3 {{\em Phys. Rev.} {\bf#1} (#2) #3.}
\def\prl#1#2#3  {{\em Phys. Rev. Lett.} {\bf#1} (#2) #3.}
\def\prs#1#2#3  {{\em Proc. Roy. Soc.} {\bf#1} (#2) #3.}
\def\ptp#1#2#3  {{\em Prog. Th. Phys.} {\bf#1} (#2) #3.}
\def\ps#1#2#3   {{\em Physica Scripta} {\bf#1} (#2) #3.}
\def\rmp#1#2#3  {{\em Rev. Mod. Phys.} {\bf#1} (#2) #3.}
\def\rpp#1#2#3  {{\em Rep. Prog. Phys.} {\bf#1} (#2) #3.}
\def\sjnp#1#2#3 {{\em Sov. J. Nucl. Phys.} {\bf#1} (#2) #3.}
\def\spj#1#2#3  {{\em Sov. Phys. JEPT} {\bf#1} (#2) #3.}
\def\spu#1#2#3  {{\em Sov. Phys.-Usp.} {\bf#1} (#2) #3.}
\def\zp#1#2#3   {{\em Zeit. Phys.} {\bf#1} (#2) #3.}

\setcounter{secnumdepth}{2} 

   
\title{TWO TOPICS IN PARTICLE-ANTIPARTICLE MIXING}

\firstauthors{T.E. Browder}

\firstaddress{Department of Physics and Astronomy, University of
Hawaii, 2505 Correa Road,
Honolulu, HI 96822, U.S.A}

\twocolumn[\maketitle]

\section{Introduction}

This paper discusses two experimental issues
in the study of particle antiparticle mixing.
We propose a new
method to extract the ratio $|V_{ts}/V_{td}|^2$
from a measurement of $\Delta \Gamma/\Gamma$ for the $B_s$ meson.
This method is experimentally
more sensitive than the conventional method for
large values of $|V_{ts}|$ but depends on the
accuracy of parton level calculations.
We then briefly discuss the implications of large CP violation
and final state interactions (FSI) in the 
experimental search for $D^0-\bar{D}^0$
mixing.

\section{A New Method for Determining $|V_{td}|/V_{ts}|^2$.}

The measurement of the mixing parameter $x_s={{\Delta m}/{\Gamma}}$ for
the $B_s$ meson is one of the goals of high energy collider experiments
and experiments planned for the facilities of the future.
A measurement of $x_s$ combined with a determination of $x_d$
the corresponding quantity for the $B_d$ meson allows the determination
of the ratio of the KM matrix elements ${|V_{td}|^2}/{|V_{ts}|^2}$
from the ratio\cite{Hagelin} 
\begin{equation}
{{x_s}\over {x_d}} = 
{{(m_{B_s} \eta_{QCD}^{B_s} B_{B_d} f_{B_s}^2)}\over
{(m_{B_d} \eta_{QCD}^{B_d} B_{B_s} f_{B_d}^2)}} {\tau_s \over \tau_d}
|{{V_{ts}}\over {V_{td}}}|^2 
\end{equation}
The factor which multiplies the ratio of KM matrix elements
is unity up to $SU(3)$ breaking effects and has
been estimated to be of order $1.3$ \cite{Brussels}. 
Since time integrated measurements
of $B_s$ mixing are insensitive to $x_s$ when mixing is maximal,
 one must make time dependent measurements in order
to extract this parameter.
A severe experimental difficulty
is the rapid oscillation rate of the $B_s$ meson, as recent experimental
limits indicate that $x_s > 8.4$\cite{Brussels} and theoretical fits to 
the Standard Model parameters suggest that $x_s$ lies in the
range $10-40$.

It should be noted that there is another parameter of the $B_s$ meson
which can also be measured, this is $\Delta \Gamma/\Gamma$, the difference
between the widths of the two $B_s$ eigenstates.
For $|V_{ts}|\sim 0.043$
this could lead to a value of $\Delta \Gamma/\Gamma$ of order
$10-20\%$ which is measurable at high energy experiments or
asymmetric B factories.
In parton calculations\cite{Hagelin}
\begin{equation}
\Delta\Gamma = {{-G_F^2 f_B^2 m_B m_b^2 \lambda_t^2} \over {4 \pi}}
~[1+ {4\over 3}
{{\lambda_c}\over{\lambda_t}} {{m_c^2} \over {m_b^2}} + O(m_c^4/m_b^4)]
\end{equation}
Comparing to the dispersive term, this gives
\begin{equation}
{{\Delta \Gamma_{B_s}} \over {\Delta m_{B_s}}}\approx {-3\over 2} \pi~ 
{{m_b^2}\over{m_t^2}}\times 
{ {\eta_{QCD}^{\Delta \Gamma(B_s)})}\over
{\eta_{QCD}^{\Delta M(B_s)} }} 
\end{equation}
where $m_b$, $m_t$ are the masses of the b and t quark 
respectively and terms of
order $m_c^2/m_b^2,~m_b^2/m_t^2$
are neglected\cite{Hagelin}.
The last factor in the above
expression, the ratio of QCD
corrections for $\Delta \Gamma$ and
$\Delta M$, is expected to be of order unity.
All of the above factors in $\Delta\Gamma$
have a common mass dependence of $m_b^2$
in the leading term. 
From equations (1) and (3),
the ratio ${{x_s}/ {x_d}}$ is then given by,
\begin{equation}
{{\Delta \Gamma_{B_s}} \over{\Delta m_{B_d}}} = {-3\over 2} \pi
  ~{{m_b^2}\over{m_t^2}}~
{{(m_{B_s} \eta_{QCD}^{\Delta \Gamma(B_s)} ~B f_{B_s}^2)}\over
{(m_{B_d} \eta_{QCD}^{\Delta M(B_d)} 
~B f_{B_d}^2)}} {{|V_{ts}|^2}\over {|V_{td}|^2}}
\end{equation}
We have assumed that the
lifetimes of the $B_d$ and $B_s$ mesons will have been
measured to sufficient precision to extract this ratio.
The above expression assumes unitarity since the
leading term  
which enters in $\Delta\Gamma$ is 
\begin{equation}
\lambda_u^2 + \lambda_c^2 + 
2 \lambda_u \lambda_c 
\end{equation}
which is expressed as
\begin{equation}
(\lambda_c + \lambda_u)^2 = (\lambda_t)^2 
\end{equation}
via unitarity of the KM
matrix.
In fact, all determinations of $|V_{ti}|$ which depend on virtual t quarks
necessarily rely on the assumed unitarity of the $3\times 3$ KM matrix.
The only way to obtain values of $|V_{ti}|$ free from this assumption
is through direct on-shell measurements of t decays.

Several authors have pointed out that the quantity 
$\Delta\Gamma (B_s)$ may be large since
there are intermediate final states such as $\bar{B}_s\to D_s^{(*)+} 
\bar{D_s^{(*)-}}$
accesible to both $B_s$ and $\bar{B}_s$
which have appreciable branching states\cite{Aleksan},\cite{Dunietz}.
The calculation of Aleksan, Le Yaouanc,
Oliver, Pene, and Raynal\cite{Aleksan} shows that the
parton model estimate and the calculation using exclusive final states
agree to within an accuracy of 30\%. 
Given the large experimental uncertainties
already present in the determination of the ratio $|V_{ts}/V_{td}|^2$,
this is not yet a serious limitation. 

In order to measure ${{\Delta \Gamma}/{\Gamma}}$, one must 
determine the lifetimes
of two samples of events. One possiblity is 
to use the large samples of $\bar{B}_s\to \psi\phi$ events 
and $\bar{B}_s\to D_s^{(*)+} \ell^- \nu$ events. 
The first sample may be dominated by
events in a single CP eigenstate as is the case for $B_{d,u}\to \psi K^*$.
This can be verified experimentally by measuring the polarization in
this decay. The latter sample of semileptonic decays will be an incoherent
mixture of both CP eigenstates. The measured lifetime difference
will be $\Delta \Gamma/\Gamma^2$, which can then be used to constrain
$|V_{ts}|^2/|V_{td}|^2$. Another possibility
 is to obtain $\Delta\Gamma$ by fitting the lifetime distribution
of a sample of $\bar{B}_s\to D_s^{(*)+}\ell^-\nu$ events to the sum
of two exponential distributions and allowing for the oscillatory term.

The sensitivity of the two methods can be roughly compared as follows.
The ALEPH lower limit on $x_s$ (8.4) corresponds to the lower limit
 $\Delta\Gamma/\Gamma$ $>0.033$~(3.3\%). A measurement of 
a $7\%$ lifetime difference corresponds to a central value of
$x_s= 15$ for a time dependent oscillation study.
For large values of $V_{ts}$, the method using $\Delta \Gamma$
eventually becomes more sensitive.
Good control of systematic effects from the
boost correction in $\bar{B}_s\to D_s^+ \ell^- \nu$ and the lifetime of
the background sample are required. Feasiblity studies of the technique
introduced here have begun at 
the CDF experiment\cite{incandela},\cite{shapiro}.



\section{Experimental Search for $D^0-\bar{D^0}$ Mixing}

As was recently noted by Blaylock, Seiden, and Nir\cite{Blaylock}
due to final state interaction (FSI) a term
proportional to $\Delta M ~t~ e^{-\Gamma t}$, which was previously
neglected, may appear in the rate
of wrong sign $D$ decays (when combining samples of $D^0$ and
$\bar{D}^0$ mesons)
even in the absence of CP violation. Moreover, in some extensions
of Standard Model which have large values of both $\Delta M$ 
and significant CP violation\cite{BGHP}, a similar term may arise.
Blaylock et al. have suggested that a value of 
$\Delta M$ larger than the
present experimental limit can be accomodated if one of these 
previously neglected terms 
destructively interferes with the other time dependent terms
which arise from mixing (proportional to $t^2 ~e^{-\Gamma t}$)
 and from doubly Cabibbo suppressed decays (DCSD) (proportional 
to $e^{-\Gamma t}$). 
They suggest that this may invalidate the use of existing limits from time
dependent mixing studies at fixed target 
experiments \cite{E691} ,\cite{E791}
to constrain extensions of the Standard Model.

The conclusion of recent work done 
in collaboration with S. Pakvasa\cite{Browder2} is that, at the present level
of sensitivity and with reasonable (though model dependent) values
for the phase difference $\delta$, 
the $\Delta M~t$ term which arises 
from FSI could change the observed event yield for
experiments which study the time dependence of mixing
by at most 10\%. This is not yet a 
significant systematic experimental limitation. The contribution
from the corresponding term proportional to $\Delta M~t$ due to
CP violation which arises 
in extensions of Standard Model is highly suppressed. This 
term is not observable at the present level of experimental sensitivity.
However, as emphasized by Liu\cite{liu1},\cite{liu2} 
and by Wolfenstein\cite{wolf2}, this term should not
be neglected as
experimental examination of the $D^0(t)-\bar{D}^0(t)$ distribution
may allow more sensitive searches for $D^0-\bar{D^0}$ mixing 
in the future if the CP violating phase is large.


\setcounter{secnumdepth}{0} 

\section{Acknowledgments}
The work described here was carried out in collaboration with
S. Pakvasa and was supported by the United States
Department of Energy and Tokkuri Tei.

\end{document}